\documentclass{emulateapj}

\shorttitle{T Dwarfs in the SDSS Southern Equatorial Stripe}
\shortauthors{Matsuoka et al.}

\begin{document}


\title{1 $\mu$m Excess Sources in the UKIDSS. I. Three T Dwarfs in the Sloan Ditital \\ Sky Survey Southern Equatorial Stripe}


\author{Y. Matsuoka\altaffilmark{1}, B. A. Peterson\altaffilmark{2}, K. L. Murata\altaffilmark{1}, M. Fujiwara\altaffilmark{1},
  T. Nagayama\altaffilmark{1}, T. Suenaga\altaffilmark{3},\\ 
  K. Furusawa\altaffilmark{4}, N. Miyake\altaffilmark{4}, K. Omori\altaffilmark{4}, 
  D. Suzuki\altaffilmark{5}, and K. Wada\altaffilmark{5}}


\altaffiltext{1}{Graduate School of Science, Nagoya University, Furo-cho, Chikusa-ku, Nagoya 464-8602, Japan; 
  matsuoka@a.phys.nagoya-u.ac.jp}
\altaffiltext{2}{Mount Stromlo Observatory, Research School of Astronomy and Astrophysics, 
  Australian National University, Weston Creek P.O., ACT 2611, Australia.}
\altaffiltext{3}{Department of Astronomical Sciences, Graduate University for Advanced Studies (Sokendai),
2-21-1 Osawa, Mitaka, Tokyo 181-8588, Japan.}
\altaffiltext{4}{Solar-Terrestrial Environment Laboratory, Nagoya University, Nagoya 464-8601, Japan.}
\altaffiltext{5}{Department of Earth and Space Science, Osaka University, Osaka, 560-0043, Japan.}


\begin{abstract}
We report the discovery of two field brown dwarfs, ULAS J0128$-$0041 and ULAS J0321$+$0051, and the rediscovery of 
ULAS J0226$+$0051 (IfA 0230-Z1), in the Sloan Digital Sky Survey (SDSS) southern equatorial stripe.
They are found in the course of our follow-up observation program of 1-$\mu$m excess sources in the United Kingdom 
Infrared Telescope Infrared Deep Sky Survey.
The Gemini Multi-Object Spectrographs spectra at red optical wavelengths (6500--10500 \AA) are
presented, which reveal that they are early-T dwarfs.
The classification is also supported by their optical to near-infrared colors.
It is noted that ULAS J0321$+$0051 is one of the faintest currently-known T dwarfs.
The estimated distances to the three objects are 50--110 pc, thus they are among the most distant field T dwarfs known.
Dense temporal coverage of the target fields achieved by the SDSS-II Supernova Survey allows us to perform a simple
time-series analysis, 
which leads to the finding of
significant proper motions of 150--290 
mas yr$^{-1}$ or the transverse velocities of 40--100 km s$^{-1}$ for ULAS J0128$-$0041 and ULAS J0226$+$0051.
We also find that there are no detectable, long-term (a-few-year) brightness variations above a few times 0.1 
mag for the two brown dwarfs\footnote{
Based on observations obtained at the Gemini Observatory, which is operated by the 
Association of Universities for Research in Astronomy, Inc., under a cooperative agreement 
with the NSF on behalf of the Gemini partnership: the National Science Foundation (United 
States), the Science and Technology Facilities Council (United Kingdom), the 
National Research Council (Canada), CONICYT (Chile), the Australian Research Council (Australia), 
Minist\'{e}rio da Ci\^{e}ncia e Tecnologia (Brazil) 
and Ministerio de Ciencia, Tecnolog\'{i}a e Innovaci\'{o}n Productiva (Argentina).}.
\end{abstract}


\keywords{brown dwarfs
  --- stars: individual (ULAS J0128$-$0041, ULAS J0226$+$0051, ULAS J0321$+$0051) --- stars: low-mass --- surveys}



\section{Introduction}

Brown dwarfs are low-mass substellar objects which could not ignite hydrogen fusion
in their collapsing phases.
They occupy the transition zone between stars and planets in the physical parameters
such as mass and temperature, hence their studies provide vital information on the 
formation of both populations.
However, their dim nature has prevented the discovery of brown dwarfs despite
the great efforts made by the early observers \citep[see][for a review]{kirkpatrick05}.
The presence of the population was first confirmed observationally by the discovery of 
GD 165B \citep{becklin88} and Gl 229B \citep{nakajima95}, which are now classified as
spectral types L and T, respectively.

Thanks to the advent of the wide-field surveys such as the Sloan Digital Sky Survey 
\citep[SDSS;][]{york00}, the Two Micron All Sky Survey \citep[2MASS;][]{skrutskie06}, 
the Deep Near Infrared Survey of the Southern Sky \citep[DENIS;][]{epchtein99}, 
the United Kingdom Infrared Telescope (UKIRT) Infrared Deep Sky Survey 
\citep[UKIDSS;][]{lawrence07}, and the Canada-France Brown Dwarf Survey \citep{delorme08},
brown dwarfs are now continuously being found.
About 600 L dwarfs and 200 T dwarfs are known today (compiled in the 
DwarfArchive.org\footnote{http://spider.ipac.caltech.edu/staff/davy/ARCHIVE/index.shtml}),
and the focus of the latest surveys tends to be on the discovery of the latest T and even
cooler Y dwarfs.
On the other hand, the number of known early-T dwarfs is still small, only 10--20 objects
in a spectral subclass.

Nonetheless, early-T dwarfs represent some key phenomena necessary for understanding
the whole brown dwarf population.
One such phenomenon is the so-called $J$-band brightening \citep{dahn02, knapp04, vrba04}, wherein early-T dwarfs
have higher $J$-band luminosity than earlier type objects which should have hotter temperatures.
It is interesting to note that a higher binary frequency of the L/T transition objects relative to 
early/mid-L and mid/late-T subclasses are suggested \citep{burgasser05}, which may be related to 
the cause of the $J$-band brightening \citep{liu06}.
While it is actively attempted to construct the models of their atmospheres in order to parameterize
their observational behaviors, the small number of known objects has hampered conclusive arguments
about the statistical properties of the population.

In this paper we report the discovery of two field early-T dwarfs, as well as the rediscovery of one,
in the SDSS southern equatorial stripe.
The dense temporal coverage achieved by the SDSS-II Supernova Survey allows us to constrain their transverse 
velocities and variability.
This is the first paper from our follow-up observation program of 1 $\mu$m excess sources in the UKIDSS.
We give a minimum description of the observation strategy in the following section, while the full description
is given in a companion paper (Y. Matsuoka et al. 2011, in preparation).
Throughout this paper, magnitudes are given in the AB system for the SDSS optical bands and in the Vega 
system for the UKIDSS near-IR bands.


\section{Observations}


\subsection{Target Selection \label{sec:targetselect}}

The three T dwarfs, 
ULAS J0128$-$0041 (R.A. 01$^{\rm h}$28$^{\rm m}$14$^{\rm s}$.41, decl. $-$00$^{\circ}$41$^{\rm m}$53$^{\rm s}$.5), 
ULAS J0226$+$0051 (R.A. 02$^{\rm h}$26$^{\rm m}$37$^{\rm s}$.55, decl. $+$00$^{\circ}$51$^{\rm m}$54$^{\rm s}$.4), and 
ULAS J0321$+$0051 (R.A. 03$^{\rm h}$21$^{\rm m}$22$^{\rm s}$.98, decl. $+$00$^{\circ}$51$^{\rm m}$05$^{\rm s}$.2),
are found on the UKIDSS Large Area Survey (LAS) images of the SDSS\footnote{
Funding for the SDSS and SDSS-II has been provided by the Alfred P. Sloan Foundation, 
the Participating Institutions, the National Science Foundation, the U.S. Department 
of Energy, the National Aeronautics and Space Administration, the Japanese Monbukagakusho, 
the Max Planck Society, and the Higher Education Funding Council for England. 
The SDSS Web site is http://www.sdss.org/.
The SDSS is managed by the Astrophysical Research Consortium for the Participating Institutions. 
The Participating Institutions are the American Museum of Natural History, Astrophysical Institute 
Potsdam, University of Basel, University of Cambridge, Case Western Reserve University, University 
of Chicago, Drexel University, Fermilab, the Institute for Advanced Study, the Japan Participation 
Group, Johns Hopkins University, the Joint Institute for Nuclear Astrophysics, the Kavli Institute 
for Particle Astrophysics and Cosmology, the Korean Scientist Group, the Chinese Academy of Sciences 
(LAMOST), Los Alamos National Laboratory, the Max-Planck-Institute for Astronomy (MPIA), the 
Max-Planck-Institute for Astrophysics (MPA), New Mexico State University, Ohio State University, 
University of Pittsburgh, University of Portsmouth, Princeton University, the United States Naval 
Observatory, and the University of Washington.} southern equatorial stripe, also known as 
the stripe 82, where the SDSS-II Supernova Survey has been carried out \citep{frieman08}.
The SDSS $i$, $z$ and the UKIDSS $Y$-, $J$-, $H$-, and $K$-band magnitudes of the targets are summarized
in Table \ref{tab:obj}\footnote{
The UKIDSS magnitudes are taken from the Data Release (DR) 3 for consistency with the target selection
strategy based on the DR3.
While the updated magnitudes are now available, they are not significantly different from the values 
presented here.}.
Note that the UKIDSS 2\arcsec.8 aperture is larger enough than the typical seeing of 0\arcsec.8
\citep{warren07}, so that the aperture losses have little influence on this work.
The targets have been selected in the course of our follow-up observation program of 1 $\mu$m excess sources
in the UKIDSS (Y. Matsuoka et al. 2011, in preparation).
Their extremely red $i - Y$ ($i_{\rm AB} - Y_{\rm Vega} > 5$) and blue $Y - J$ ($Y_{\rm Vega} - J_{\rm Vega} < 1$) 
colors, along with the stellar appearances, suggested that they can be either of the two rare populations, i.e.,
highest-redshift ($z > 6.5$) quasars or brown dwarfs \citep[e.g.,][]{venemans07}.
The discovery of the former objects are our ultimate goal, which has not been achieved in the previous projects
such as the SDSS quasar survey \citep[][and references therein]{fan06}, the Canada--France High-$z$ Quasar Survey
\citep[][and references therein]{willott10}, and the Tokyo-Stromlo Photometry Survey \citep{matsuoka08}.

\begin{table*}
\begin{center}
\caption{Red-optical and Near-IR Magnitudes \label{tab:obj}}
\begin{tabular}{ccccccc}
\tableline\tableline
     & $i_{\rm AB}$ & $z_{\rm AB}$ & $Y_{\rm Vega}$ & $J_{\rm Vega}$ & $H_{\rm Vega}$ & $K_{\rm Vega}$\\
Object & (mag) & (mag) & (mag) & (mag) & (mag) & (mag)\\
\tableline
ULAS J0128$-$0041 & $>$25.64 & 20.57 (0.04) & 18.48 (0.07) & 17.62 (0.05) & 16.91 (0.04) & 16.52 (0.06)\\
ULAS J0226$+$0051\tablenotemark{a} & $>$25.88 & 21.32 (0.08) & 19.16 (0.11) & 18.33 (0.08) & 17.83 (0.12) & 17.69 (0.16)\\
ULAS J0321$+$0051 & $>$25.84 & 22.23 (0.17) & 20.06 (0.18) & 19.19 (0.10) & 18.76 (0.20) & 18.74 (0.26)\\
\tableline
\end{tabular}
\tablecomments{Given above are the PSF magnitudes for the SDSS $i$ and $z$ bands and the 2\arcsec.8 aperture
  magnitudes for the UKIDSS $Y$, $J$, $H$, and $K$ bands. Values in the parentheses represent 1$\sigma$ errors,
  while 2$\sigma$ upper limits are given for the non-detected sources.
  The magnitudes are extracted from the SDSS DR 7 and the UKIDSS DR 3.\\
  $^{\rm a}$This object is identical to IfA 0230-Z1 ($J$ = 18.17 $\pm$ 0.03 mag and $H$ = 17.83 $\pm$ 0.04 in the
  MKO system) discovered by \citet{liu02}.}
\end{center}
\end{table*}

We carried out the follow-up photometry of a few dozen 1 $\mu$m excess sources, including the present three objects,
before the final spectroscopy.
The optical imaging observations were carried out during 2009 August--September, with a special $i$-band filter with the 
blueward transmission extending to 6300 \AA\ (at the half of the maximum transmission) installed to the Imager 
mounted on the Australian National University (ANU) 2.3-m Advanced Technology Telescope at the Siding Spring Observatory.
At the near-IR wavelengths, the SIRIUS camera \citep{nagashima99, nagayama03} of the Infrared Survey Facility (IRSF) 
1.4 m telescope at Sutherland, South African Astronomical Observatory was used for the follow-up observations.
They were conducted in the two periods, 2009 June--July and September--October.
The magnitudes listed in Table \ref{tab:obj} are found to be robust in these follow-up photometry.
The 1.8 m Microlensing Observations in Astrophysics \citep[MOA;][]{bond01,sumi03} II telescope is also 
used for our project, although the present three targets are not covered by the MOA observations.


\subsection{Spectroscopy}
Based on the revised photometry, we selected 15 objects with the strongest 1-$\mu$m excesses and obtained their spectra.
The three T dwarfs found in the above targets are the subject of this paper, while the rest of them appear to be other 
classes of objects.
The results of the whole observations will be presented in a next paper.
The spectroscopy was carried out using the Gemini Multi-Object Spectrographs \citep[GMOS;][]{hook04} mounted on the Gemini 
North telescope (Program ID: GN-2010B-Q-102).
The observation journal is given in Table \ref{tab:obs}.
The R400-G5305 grating with the central wavelength set to 8500 \AA\ was used with the blocking filter RG610-G0307, 
so that the wavelength range from 6500 \AA\ to 10500 \AA\ was covered.
Since the targets are too faint to view on the acquisition images, we adopt either of the following acquisition techniques.
For ULAS J0226$+$0051, the "blind offset" acquisition was performed using a nearby bright star as a reference source.
For the other two targets without suitable nearby stars, we acquired the bright stars within a few arcminutes at the slit center 
and configured the instrument position angles so that the targets fall off-center of the slit.
In order to avoid a large flux loss due to misalignment of the slit, which may be caused by the blind acquisitions,
we adopted the relatively wide slit of 1\arcsec.5 in width.
It results in the moderate wavelength resolution of $R \sim 600$, which is still enough for identifying the objects.
The single exposure time was 900 s for all the science targets.
The targets were offset by 10\arcsec\ along the spatial axis of the slit between the exposures, which enables
good subtraction of the sky background with the prominent fringe patterns at the red part of the spectra.

\begin{table*}
\begin{center}
\caption{Observation Journal of Gemini/GMOS Spectroscopy\label{tab:obs}}
\begin{tabular}{cccc}
\tableline\tableline
Object & Date & Exp. Time (s) & Type\\
\tableline
ULAS J0128$-$0041 & 2010 Aug. 30 & 1800  & Science target\\
ULAS J0226$+$0051 & 2010 Sep. 19 & 3600  & Science target\\
EG 131            & 2010 Sep. 20 & 120   & Standard star\\
ULAS J0321$+$0051 & 2010 Sep. 22 & 10800 & Science target\\
\tableline
\end{tabular}
\end{center}
\end{table*}

The data were reduced in a standard manner using the Gemini IRAF\footnote{IRAF is distributed by the 
National Optical Astronomy Observatory, which is operated by the Association of Universities for Research in Astronomy, Inc.,
under cooperative agreement with the National Science Foundation.} package, version 1.10.
First the bias subtraction and flat fielding were performed, then the CuAr spectra were used to rectify and give the 
wavelength solutions to the images.
The fringe patterns were successfully eliminated by subtracting the adjacent exposures.
Then the residual sky background was estimated from the fluxes of the nearby pixels in the spatial direction and removed.
The instrument sensitivity was calibrated with the observed spectrum of the spectroscopic standard star EG 131.
We did not observe a telluric standard star on each night, hence the accurate correction for the atmospheric absorption 
is not possible.

\begin{figure*}
\epsscale{0.7}
\plotone{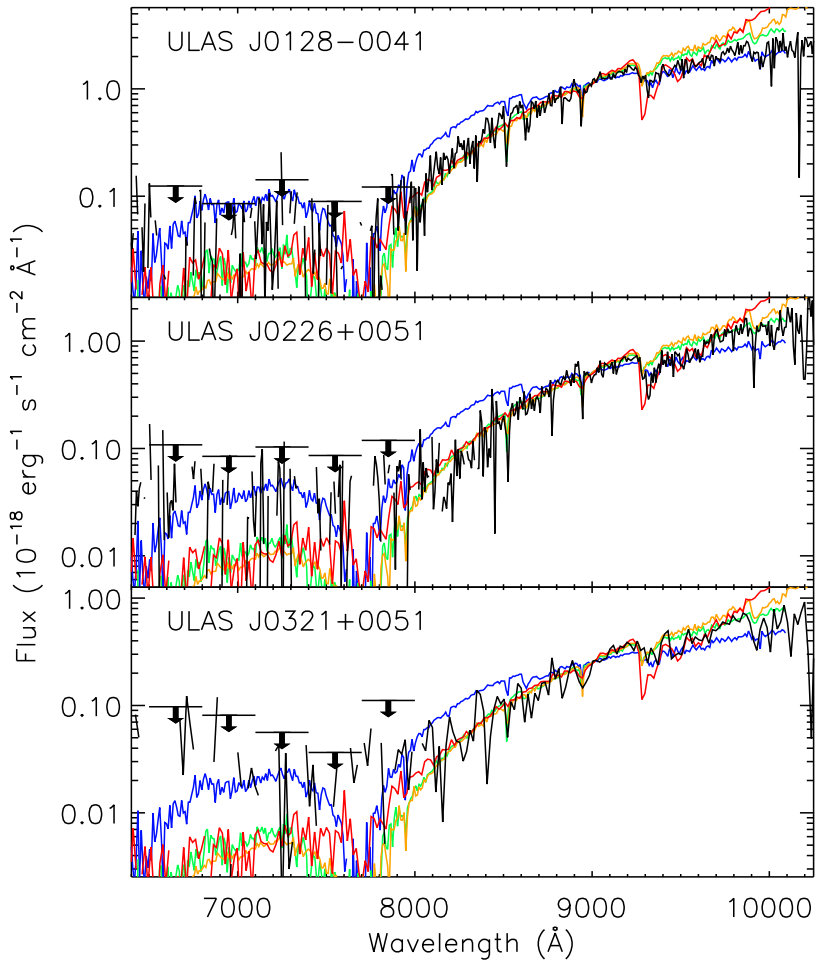}
\caption{Gemini/GMOS spectra of the three T dwarfs (black).
  The horizontal lines with the arrows at $\lambda <$ 8000 \AA\ represent the broadband 2$\sigma$ upper limits of the
  observed fluxes.
  The template spectra of the L8 (blue), T2 (green), T5 (orange), and T8 (red) dwarfs, normalized
  at $\lambda$ = 9000 \AA, are also plotted for comparison.\label{spec}}
\end{figure*}

We show the reduced spectra in Figure \ref{spec}.
They are not detected above 3$\sigma$ significance at the wavelengths $\lambda <$ 8000 \AA, where we give the broad-band 
2$\sigma$ upper limits calculated in the wavelength intervals $\lambda$ = 6500--6800, 6800--7100, 7100--7400, 7400--7700, 
and 7700--8000 \AA.
Note that the H$_2$O absorption bands at $\lambda \sim$ 9300 \AA\ are contaminated with the telluric absorptions.

\section{Analysis}

\subsection{Spectral Type}

Since the signal-to-noise ratios of our spectra are not very high, we determine the spectral types of the targets
with their broadband spectral energy distributions (SEDs) rather than with the individual absorption features.
The extremely-red SEDs suggest that they are the members of brown dwarfs with the spectral types later than L.
In Figure \ref{spec} we compare the observed spectra with those of the L8, T2, T5, and T8 dwarfs which are commonly used 
as the reference objects (anchor points) for the classification at the red optical wavelengths \citep{kirkpatrick99,burgasser03}, 
namely, 2MASSW J1632291$+$190441 (L8), SDSSp J125453.90$-$012247.4 (T2), 2MASSI J0559191$-$140448 (T5), and 
2MASSI J0415195$-$093506 (T8).
The three targets are clearly redder than the L8 dwarf and show the overall agreements with the T-dwarf spectra, while the 
spectral slopes at the longest wavelengths ($>$ 9800 \AA) indicate that they are not in late-T subclasses.

\begin{figure}
\epsscale{1.0}
\plotone{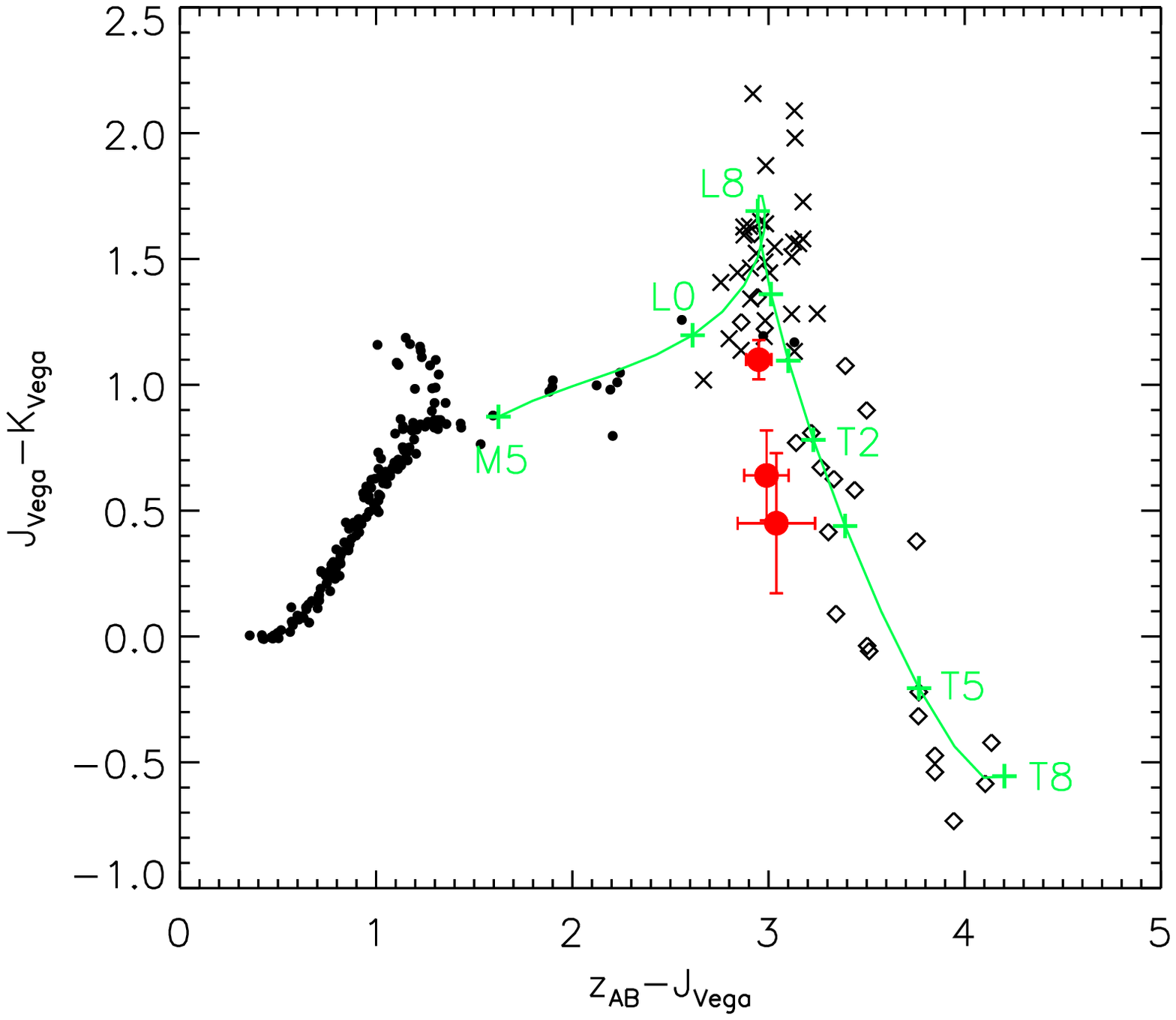}
\caption{$z - J$ vs. $J - K$ two-color diagram of stars and brown dwarfs.
  The small circles represent O--M stars while the crosses and the diamonds represent L and T dwarfs, respectively, 
  taken from \citet{hewett06}.
  The empirical mean relation between the colors of M5--T8 dwarfs \citep{kakazu10} are also plotted (green line)
  with the plus marks placed at the positions of M5, L0, L8, T0, T1, T2, T3, T5, and T8.
  Our targets are shown by the red circles with error bars (ULAS J0128$-$0041, ULAS J0226$+$0051, and ULAS J0321$+$0051,
  from top to bottom).\label{spec_type_nir}}
\end{figure}

The classification of them as early-T type is also supported by their optical to near-IR colors.
We show the $z - J$ versus $J - K$ two-color diagram of stars and brown dwarfs in Figure \ref{spec_type_nir}.
The colors of O--M stars and L, T dwarfs are obtained from the compilation by \citet{hewett06}, while the empirical 
mean relation between the colors of M--T dwarfs is taken from \citet{kakazu10}.
Note that all the near-IR magnitudes are consistently based on the Mauna Kea Observatories (MKO) system \citep{tokunaga02},
which is important because different photometric systems can lead to the different near-IR magnitudes by up to 0.4 mag
for T dwarfs \citep{stephens04}.
The three targets are located in the region occupied by the early-T dwarfs, and the most plausible classification
would be T1 for ULAS J0128$-$0041, T2 for ULAS J0226$+$0051, and T3 for ULAS J0321$+$0051 (summarized in Table \ref{tab:props}).
However, we emphasize that the above typing is not meant to be conclusive: the uncertainty of $\pm$1 subclass seems plausible.
We note that ULAS J0321$+$0051 is one of the faintest T dwarfs known.

\begin{table*}
\begin{center}
\caption{Estimated Properties \label{tab:props}}
\begin{tabular}{ccccc}
\tableline\tableline
                  & Spectral & Distance & Proper Motion   & Transverse Velocity\\
Object            & Type     & (pc)     & (mas yr$^{-1}$) & (km s$^{-1}$)\\
\tableline
ULAS J0128$-$0041 & Early-T (T1) &  50 $\pm$ 10 & 150 $\pm$ 30  & 40 $\pm$ 10\\
ULAS J0226$+$0051 & Early-T (T2) &  70 $\pm$ 20 & 290 $\pm$ 140 & 100 $\pm$ 50\\
ULAS J0321$+$0051 & Early-T (T3) & 110 $\pm$ 30 & $\cdot\cdot\cdot$ & $\cdot\cdot\cdot$\\
\tableline
\end{tabular}
\tablecomments{Spectral subclasses given in the parenthesis are not conclusive. See the text.}
\end{center}
\end{table*}

While ULAS J0128$-$0041 and ULAS J0321$+$0051 are the newly discovered T dwarfs, ULAS J0226$+$0051 has 
previously been reported by \citet{liu02}.
The object, named IfA 0230$-$Z1 in the discovery paper, was identified in the Institute for Astronomy (IfA)
Deep Survey conducted at the University of Hawai'i.
\citet{liu02} present a $H$-band spectrum as well as $I$, $z$, $J$, and $H$-band magnitudes of the object, 
from which they estimate a spectral type of T3--T4.
Later the updated near-IR classification of T3 is given by \citet{burgasser06}.
Our spectral typing is consistent with these near-IR classifications considering the typical 
uncertainty of one spectral subclass.

\subsection{Distance, Proper Motion, and Variability}

We estimate the distances to the three T dwarfs by using the luminosity--spectral type relation provided by \citet{liu06}.
The $H$-band magnitudes are used for this purpose, since $H$-band luminosity of early-T dwarfs is relatively independent
of the spectral subclasses which are not precisely known in the present case.
Assuming the $H$-band absolute magnitude of 13.5 $\pm$ 0.5 mag \citep{liu06} for T1--3 spectral types, the distance 
estimates to ULAS J0128$-$0041, 
ULAS J0226$+$0051, and ULAS J0321$+$0051 are 50 $\pm$ 10 pc, 70 $\pm$ 20 pc, and 110 $\pm$ 30 pc, respectively.
The lower limit of the estimate for ULAS J0226$+$0051 is consistent with 49 $\pm$ 9 pc presented by \citet{liu02}.
As compiled by \citet{kakazu10}, there are only a handful of spectroscopically confirmed T dwarfs known today at distances
beyond 60 pc, and only a few beyond 100 pc.
Therefore the objects presented here are among the most distant T dwarfs.

The three dwarfs are found in the SDSS southern equatorial stripe, where the intensive repeat scans were carried out 
in the SDSS-II Supernova Survey.
The dense temporal coverage over several years allows us a simple time-series analysis of the objects.
We retrieve all the $z$-band images of the target fields from the SDSS DR 7 archive \citep{abazajian09} and create 
the stacked images of each year.
The stacking is not performed when less than five images are available in a year.
It results in the deep images of the years 2002, 2005, 2006, 2007 for ULAS J0128$-$0041, years 2002, 2003, 2005, 2006, 2007
for ULAS J0226$+$0051, and years 2002, 2005, 2006, 2007 for ULAS J0321$+$0051.
We measure the coordinates and magnitudes of the objects with the {\it Source Extractor}, version 2.5 \citep{bertin96}, if
the targets are detected on the stacked images (more than four adjacent pixels above 1.5$\sigma$ of the local background are 
defined as detections).
The results for the detected sources are summarized in Table \ref{tab:timeseries}.
Unfortunately, ULAS J0321$+$0051 is too faint to be detected on any of the year-based stacked images.
The coordinates are shown as the relative offsets (milliarcsecond; mas) from the UKIDSS positions.
The magnitudes are measured in 3\arcsec.0 apertures and are corrected so that the mean measured magnitudes match the 
point-spread function (PSF)
magnitudes listed in Table \ref{tab:obj}, whose differences represent the mean aperture losses ($-$0.25 and $-$0.20 mag 
are added to the measured magnitudes of ULAS J0128$-$0041 and ULAS J0226$+$0051, respectively).

The positional information of ULAS J0226$+$0051 is also available from \citet{liu02}, who report the coordinate obtained
on the epoch of 2001 October with the Suprime-Cam mounted on the Subaru telescope.
Its relative offset from the UKIDSS position is $\Delta$R.A. = $+$0--$+$1300 mas and $\Delta$decl. = $+$300 mas
(since \citet{liu02} round the R.A. to first decimal place in the paper, we assume it to be 
02$^{\rm h}$26$^{\rm m}$37$^{\rm s}$.55--37$^{\rm s}$.64).
The derived $\Delta$decl. value, on the year 2001, is clearly inconsistent with the increasing declination of this object 
indicated in Table \ref{tab:timeseries}.
The major reason for this inconsistency (though within an arcsecond) is likely the different astrometric calibrations adopted,
therefore we do not use the Suprime-Cam coordinate in the following arguments.

Considering the typical SDSS astrometric uncertainty of 100 mas at the limiting depth \citep{pier03}, significant spatial 
movements are detected for the two dwarfs in Table \ref{tab:timeseries}.
We derive the proper motions from the two measurements with the longest baselines, i.e., 2002--2007 for 
ULAS J0128$-$0041 and 2006--2007 for ULAS J0226$+$0051.
The transverse velocities are then calculated using the distances estimated above.
We list the results in Table \ref{tab:props}.
These kinematic properties provide vital information not only on the structure of the Galaxy but also on the ages of the
substellar systems when combined with the appropriate models.
The derived transverse velocity of ULAS J0128$-$0041 is in agreement with the median value $\sim$ 30 km s$^{-1}$ of early-T 
dwarfs found by \citet{faherty09}, suggesting that the object is a member of typical field T dwarfs.
On the other hand, ULAS J0226$+$0051 may have higher velocity than the typical objects, suggesting that this dwarf may be
an older member of the Galactic thick disk or halo population \citep[e.g.,][]{nordstrom04}.


\begin{table}
\begin{center}
\caption{Time-series Analysis \label{tab:timeseries}}
\begin{tabular}{ccccc}
\tableline\tableline
                  &          & $\Delta$R.A. & $\Delta$decl. & $z_{\rm AB}$ \\
Object            & Year     & (mas)        & (mas)         & (mag)       \\
\tableline
ULAS J0128$-$0041 & 2002 & $+$320 & $+$170 & 20.57 (0.15)  \\ 
                  & 2005 & $+$20  & $-$40  & 20.55 (0.09)  \\ 
                  & 2006 & $-$60  & $-$30 & 20.59 (0.09)  \\ 
                  & 2007 & $-$350 & $-$210 & 20.55 (0.08)  \\ 
\hline
ULAS J0226$+$0051 & 2006 & $+$190 & $-$200 & 21.32 (0.14) \\ 
                  & 2007 & $+$350 & $+$40  & 21.32 (0.15) \\ 
\tableline
\end{tabular}
\tablecomments{Coordinates are given as the relative offsets from the UKIDSS positions. 
  Values in the parenthesis represent 1$\sigma$ photometry errors.}
\end{center}
\end{table}

As for the source brightness, we find no detectable long-term variations in the $z$ band.
The Suprime-Cam $z$-band photometry gives $z_{\rm AB}$ = 21.44 $\pm$ 0.15 mag for ULAS J0226$+$0051 \citep{liu02} when
the Vega-to-AB magnitude conversion of $+$0.533 mag \citep{hewett06} is used, which is also in agreement with
those listed in Table \ref{tab:timeseries}.
Note that the Suprime-Cam $z$-band filter is very similar to that of the SDSS.
There have been several observational searches for short-term (minutes to weeks) variability in brown dwarfs at the L/T 
transition \citep[e.g.,][]{enoch03, clarke08, artigau09}, which could give a stringent constraint on the models of their 
atmospheres.
Here our inquiry into the long-term (years) variations is not based on a model prediction, but rather motivated by a need 
for the {\it empirical} constraints on the stability of brown-dwarf colors, which are essential in defining the selection 
strategies 
of high-redshift quasars or brown dwarfs from photometry data obtained on different dates (see Section \ref{sec:targetselect}).
The lack of significant variations over a few times 0.1 mag may indicate that the variability of early-T dwarfs do not
cause significant uncertainty in the selection strategies, although more samples, including other subclasses of brown
dwarfs and time-series near-IR data, are needed for a firm conclusion.

\section{Summary}

This is the first paper from our follow-up observation program of 1 $\mu$m excess sources in the UKIDSS.
In this paper we present the newly discovered two brown dwarfs, ULAS J0128$-$0041 and ULAS J0321$+$0051, 
as well as the re-discovered ULAS J0226$+$0051 (IfA 0230-Z1), in the SDSS southern equatorial stripe.
The follow-up imaging observations were carried out with the optical Imager on the ANU 2.3-m telescope and
the near-IR SIRIUS camera on the IRSF telescope.
Then we obtained their red optical (6500--10500 \AA) spectra with the GMOS on the Gemini North telescope 
in order to identify the targets.
The spectra reveal that the objects belong to early-T dwarf subclasses, which is
also supported by their optical to near-IR colors.
ULAS J0321$+$0051 turns out to be one of the faintest currently-known T dwarfs.

We estimate the distances to the dwarfs to be 50--110 pc from the empirical luminosity--spectral type relation of
brown dwarfs.
By taking advantage of the dense temporal coverage of the target fields achieved by the SDSS-II Supernova Survey, 
we create the stacked images of each year during 2002--2007 and conduct a simple time-series analysis.
As a result, we find the significant proper motions of 150 and 290 mas yr$^{-1}$, which are converted to
the transverse velocities of 40 and 100 km s$^{-1}$, for ULAS J0128$-$0041 and ULAS J0226$+$0051, respectively.
It suggests that ULAS J0128$-$0041 is a member of typical field T dwarfs, while ULAS J0226$+$0051 may have the higher 
velocity than average and be an older member of the Galactic thick disk or halo population.
We also look into the possible long-term (a-few-year) brightness variations of the two objects, and find
no detectable variations above a few times 0.1 mag.
This suggests that the variability of early-T dwarfs does not cause significant uncertainty in the selection 
strategies of high-redshift quasars or brown dwarfs with photometry data taken on different dates, although more 
observations are needed to reach a firm conclusion.

\acknowledgments

We are grateful to the referee, Sandy Leggett, for giving very useful comments and suggestions.
We thank the staff of the Siding Spring Observatory, the South African Astronomical Observatory, and the Gemini 
Observatory (including Alexander Fritz) for the support during the observations. The IRSF team at Nagoya University, 
Kyoto University, and 
the National Astronomical Observatory of Japan has provided a great help for the IRSF observations.
This work was supported by Grants-in-Aid for Scientific Research (21840027, 22684005) and the Global COE Program 
of Nagoya University "Quest for Fundamental Principles in the Universe" from JSPS and MEXT of Japan.






\clearpage


\end{document}